# Development of a Home Automation System Using Wireless Sensor/Actuator Nodes

Babatope S. Ayo

**Abstract** — *This work presents the design and implementation of a wireless home monitoring and automation system consisting of wireless sensor/actuator nodes, wireless camera, and a home server. The low-cost wireless sensor/actuator node features temperature, light intensity and motion sensors, and actuator driver circuits for the control of motors, heaters, and lights. Server and client programs used to monitor and control the home were also developed. The home server receives and processes sensor readings, such as temperature and light intensity readings, and also transmits user commands to wireless nodes. The system provides ambient condition monitoring, graphing of sensor data, intrusion detection, automated device control, and video monitoring in order to achieve improved security and comfort in the home. In addition, users have the flexibility of determining sensor-actuator interaction at run-time. The developed system could also put the home in various configurable modes based on user requests, time or environmental cues[1].*

## I. INTRODUCTION

Over the years, the sizes and costs of processors and wireless transceivers have been shrinking tremendously. This has led to the proliferation of smart devices with embedded processing and communication capabilities, such as smartphones and wireless sensors. One area of application that has attracted a lot of attention is the smart home, in which groups of such devices form wireless home automation networks (WHANs).

WHANs refer to networks of sensors, actuators and other devices that interact over wireless links to provide services which improve the safety, security or comfort of homes and their occupants [1]-[3]. Sensors measure parameters of interest, such as ambient or health variables. Ambient conditions include temperature, humidity or light intensity. Health parameters monitored by sensors include electrocardiogram (ECG), weight, blood pressure, and heart rate [4]-[5]. Actuators, on the other hand, execute desired activities in the home, such as switching of lights or heaters.

Interfaces serve as the platform through which users can control or interact with the home devices and environment. Examples of such interfaces are mobile phones, custom-built touchscreen consoles, and computer software. Services provided by WHANs include environmental monitoring, intrusion detection, health monitoring, energy management, remote/automated device control, and detection of hazardous conditions such as gas leaks [6]-[9].

A lot of wireless protocols have been implemented in the home scenario. Some of these standards include IEEE 802.15.4, ZigBee, 6LoWPAN, Bluetooth, and IEEE 802.11 [1]. They operate mostly in the license-free Industrial, Scientific and Medical (ISM) bands of 2.4 GHz and sub-1 GHz. Some of these, such as Bluetooth and IEEE 802.11, were not originally designed for such low power applications but have gained popularity because of the existence of well-established consumer products using the standards. When compared with wired systems, WHANs have the advantages of easy installation and configuration, better aesthetics, and easy troubleshooting/repair because of their wireless and ad-hoc nature.

This work discusses the development of a wireless home automation network using wireless nodes and a camera operating in the 2.4 GHz radio-frequency (RF) band. In addition, embedded and Java GUI software were developed to enable the provision of services in the home, such as ambient condition monitoring, automated device control and intrusion detection, in order to improve living conditions and safety in the home.

## II. RELATED WORK

A lot of work has been done in the area of applying wireless sensors to home monitoring and automation, and the development of relevant interfaces [3]. A short review of some of the research follows.

Yahaya *et al* [10] designed a web-based monitoring system that allows the continuous monitoring of room temperature over the internet and X10. Through a speech synthesizer, it can verbally inform users of current temperature. The system has limited use since it focussed only on temperature measurements.

Ahn *et al* [11] designed a home automation system that features a personal digital assistant (PDA)-enabled mobile robot. Home appliances are connected to a gateway through wired or wireless links. Whenever the home network is faulty or visual information is needed, the robot is commanded by a server program to go close to the specified device and control it. The robot also provides visual feedback so that users can confirm the state of the controlled device.

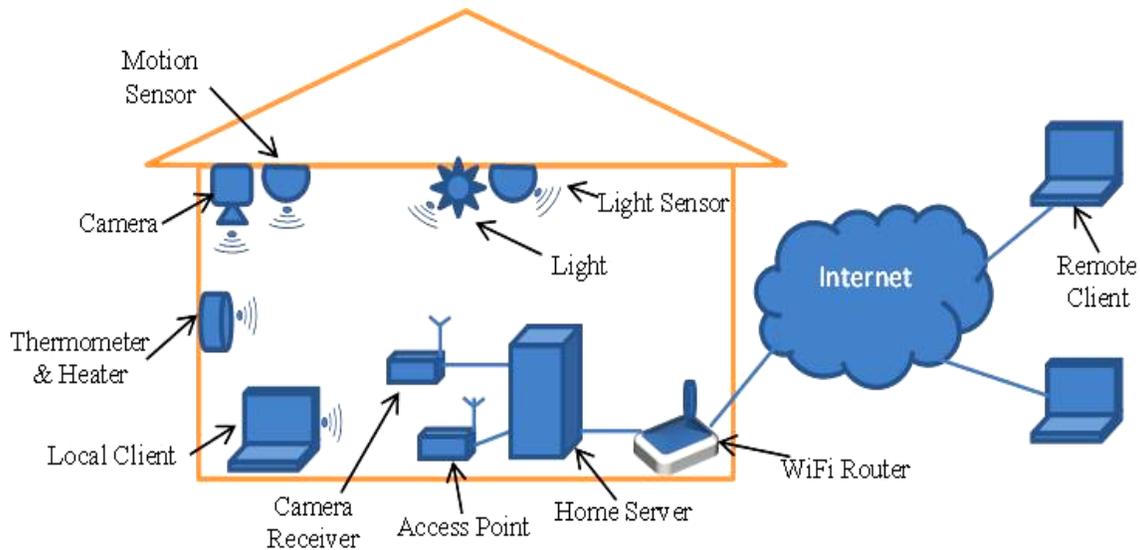

Fig. 1. Wireless home automation design architecture.

Zhang *et al* [12] developed ZigBee wireless sensor and actuator nodes for home monitoring and appliance control system. However, the system did not provide support for data storage and video monitoring. These additional features are desirable in order to improve system capabilities and user experience.

Therefore, this work presents a simplified and cost-effective internet-enabled home monitoring system with wireless sensors, actuators, and camera with many important features such as ambient condition monitoring, video monitoring, data storage, and automated device control.

## III. SYSTEM ARCHITECTURE

The designed WHAN is made up of wireless end devices (EDs) and cameras distributed around the house, the home server, and clients that access the home devices over the internet or home network (Fig. 1). The EDs measure ambient conditions and relay the information to the access point (AP). They also receive commands from the AP to control devices. The AP communicates information to and from the ED to the home server. The EDs and AP form the wireless sensor/actuator network of the home. The WHAN has an operating frequency of 2.4 GHz, a band that is license-free worldwide. This was chosen so that the system could be deployed in most places with little or no modifications.

### A. Home Server

The home server acts as the final aggregation and processing platform for sensor and visual data. It functions as the central controller and monitor of the home environment and devices. It also stores and retrieves data, logs and settings relevant for the operation of the system. It runs the graphical user interface (GUI) software that implements the above functionalities. In addition, it features server components that accept clients' requests from user devices. The home server serves as the gateway between the WHAN and home wired/wireless local area network (LAN). It was implemented using a laptop computer. It can also be implemented on an existing home/media server and so eliminate extra costs of a dedicated gateway device.

### B. Clients

These are end devices, such as computers, which home occupants use to monitor and control the home. They can be local and remote. Local clients are part of the wired/wireless LAN and access the WHAN directly through the home server. Remote clients, on the other hand, access the home over the internet. Clients run software that directs the users' intentions and requests to the home server which, in turn, communicates with the WHAN.

### C. Access Point

The AP processes the information from the EDs and transforms it to the format required by the home server. It also forwards user commands to each home device on the WHAN. It communicates with the home server using the USB virtual serial port. A wireless node connected to the home server through a USB debugger was used as the AP in this work. The AP and ED communicate using a proprietary low power RF protocol.

### D. End Devices

EDs feature sensors such as temperature, motion, and light sensors. They also have driver circuits for controlling home appliances. They communicate with the access point (AP) using the wireless radio transceiver. A detailed functional diagram of an ED is shown in Fig. 2. The ED is composed of a microcontroller for processing functions, a radio transceiver for communication capabilities, sensors for condition monitoring, driver circuits for device control, and a power supply.

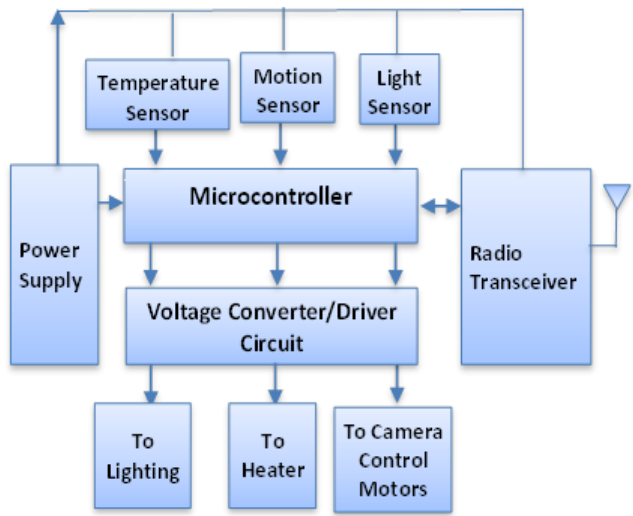

**Fig. 2.** Functional diagram of end device.

The wireless transceiver and microcontroller components were implemented using a wireless module. The other components such as sensors and actuator driver circuits were developed on an expansion board. This board was connected to the expansion pins of the wireless target board. Temperature, light intensity and PIR (pyroelectric infrared) motion sensors were implemented. Drivers for heater, lighting and camera control motors were also on the board. Voltage converters, buffers and drivers for the motors were implemented using logic inverter integrated circuits (ICs) and Darlington arrays. The inverters acts as buffers to protect the microcontroller in the case that excessive current is drawn by the driver ICs and also interface the 3 V microcontroller outputs to the 5 V levels of the driver inputs [3]. Another buffer/driver was used to control the relay that switches the heater. Fig. 3 shows the developed printed circuit board (PCB) for the expansion board.

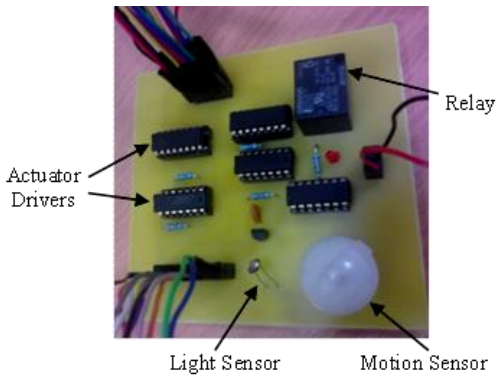

**Fig. 3.** Wireless node expansion board.

### E. Wireless Camera

This provides the visual monitoring of the home environment in which it is placed. The camera operates on a frequency of 2.4 GHz. A mechanism was constructed to handle the panning and tilting of the camera using an ED (Fig. 4). This allows the remote control of camera position for better viewing experience. Camera control signals from the AP are received by the ED that then drives the required DC stepper motors. The camera could be panned 360º and tilted over a total angle of 303º. However, dedicated transmitter and receiver were used for video data because of the low data rate and power reserve of the ED.

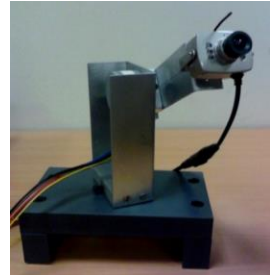

**Fig. 4.** Wireless camera and pan/tilt mechanism.

## IV. HOME SERVER SOFTWARE

The Home Server program receives sensor and visual data from the AP and wireless camera receiver through serial/USB communication. It processes and stores sensor data, which it makes available to the clients. It receives client commands and passes the relevant ones to the wireless nodes through the AP. In addition, it receives and displays video data from the wireless camera, as required by the user. It also provides client devices access to the camera feed and control. Its architecture is shown in Fig. 5. The software was written in Java, a well-established free programming language. Major components of the program are discussed as follows.

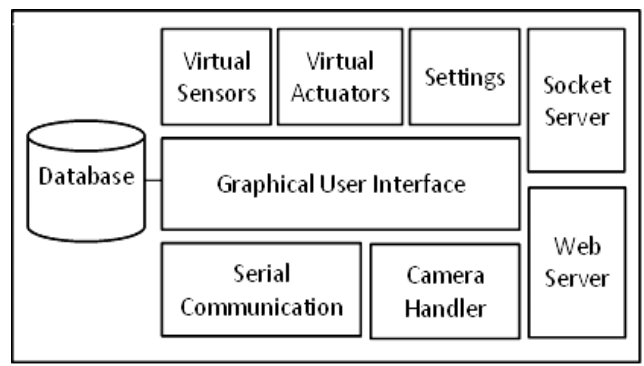

**Fig. 5.** Components of the home server software.

### A. Serial Communication Module

This software component handles communication with the AP over the virtual serial port. It receives sensor data from the AP and sends control messages to EDs based on user commands.

*B. Home Server GUI*

This provides a visual platform for the user to monitor and control the home and its devices. It provides the usual visual controls such as buttons and textboxes. It also alerts users whenever events, such as intrusion detection, occur.

*C. Device Class*

This is a class of logical home devices. It is an abstraction of physical devices, such as temperature sensors or lights. It has properties such as name, state, type, value and set point. These properties were obtained by observing the general characteristics of physical devices and then representing them in software. Various methods were used to access and modify these attributes. Devices were grouped into virtual sensors and actuators.

*D. Camera Handler*

This component handles the access, processing, display and streaming of video data to users. It receives camera feed through the camera receiver, transposes it, and then streams it so that client devices can access it. In addition, it transmits commands used to tilt/pan camera to the ED attached to the camera.

*E. Database*

A relational database was built for the system. This database contains details used to authenticate users, system mode settings, logs of important events and sensor readings data.

*F. Socket Server*

This component handles communication between the home server and clients. The home server sends sensor readings and device status over this link. It also receives client control commands over this link. At start up, the transmission control protocol (TCP) socket server is initialized and listens for connections from clients on a given port. For each successful connection, it creates a thread to handle communication to and from the client [13].

*G. Web Server*

This stand-alone component is used to deploy client applications using Java Web Start to client computers [13]. Users merely point their browsers to the web link of the home server to install the client application, which can be used to monitor and control the home from a remote location.

*H. Settings*

With the settings component, the user can configure the various parameters of each home device such as set point, switching times and low/high threshold alarm values. The user can customize sensor-actuator interaction during run-time. For example, the user can set lighting to come on when light sensor readings fall below a particular level, which is often the case at night-time.

## V. CLIENT SOFTWARE

The client software is used to access the home environment from a remote location. The client software could be run on user devices on the local computer network or over the internet. The program consists of the GUI, client socket component, camera handler, device class, settings menu and sensor data graphs. Fig. 6 (a) shows a diagrammatic representation of the client software components. A screenshot can be seen in Fig 6 (b).

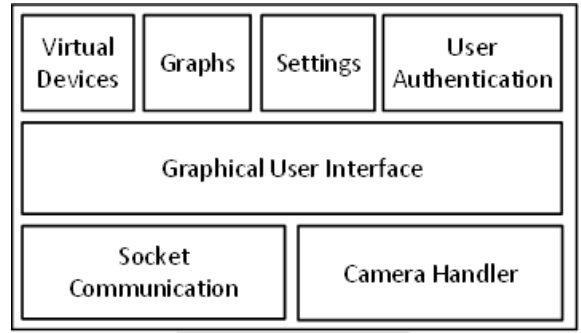

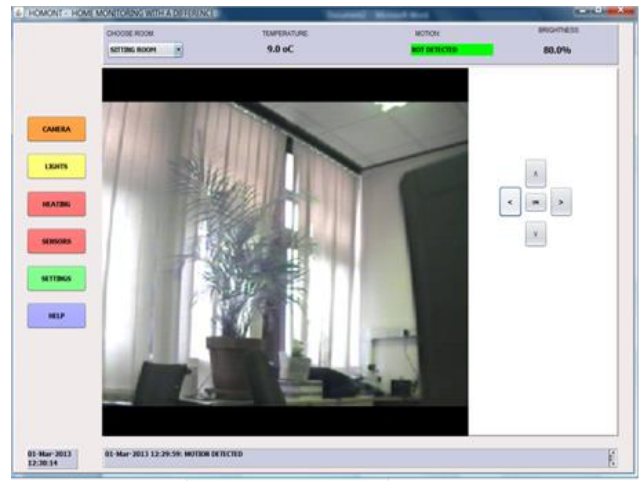

Fig. 6. Developed client software.

The program connects to the home server through a web socket. Through this link, it receives sensor data readings/device statuses and sends user commands to devices. The client software also connects to the video stream from the home server using the camera module. Before a client is allowed to access the home environment, the authentication data (username and password) provided by the user is verified. The sensor data charts display sensor values graphically for more intuitive observation and analysis by the user. The client settings component is similar to that of the server software and allows users to configure properties of devices.

## VI. FEATURES OF THE SYSTEM

A detailed presentation of the features provided by the developed system is given below.

### A. Ambient Home Condition Monitoring and Regulation

The system is able to measure parameters of the home environment such as temperature and light intensity. The system can also be used to maintain these parameters at desired levels. In addition, alarms are triggered if the measured parameters exceed low or high threshold values.

### B. Graphing of Sensor Data

Sensor data may provide valuable insights to home conditions. For example, an abnormal change in room temperature characteristics may be due to a faulty boiler. An intuitive way of analyzing these data is through graphs, which make it easier to discover trends. The user interface of this system displays sensor readings as a function of time.

### C. Intrusion Detection

With the use of PIR motion sensors, intrusion into the home is detected and alerts sent to the appropriate persons and clients.

### D. Device Control

With the aid of the GUI, home lighting and heating can be switched by the user remotely. Device control could also be achieved using timers that switch devices on and off at specified times. Another form of automated control is based on sensor readings, such as from a motion sensor (when someone is in the room) or a light sensor (for example, at night time). The heater could be switched on/off automatically in order to maintain a given temperature.

### E. Video Monitoring

With this system, parts of the home can be visually monitored from other parts of the home or over the web. In addition, the camera could be panned or tilted for proper viewing. The user is provided with an experience that is close to being present in the home physically.

### F. Modes

The home can be put in different modes depending on the prevailing circumstances of the home or its occupants. For example, when it is night time as inferred from the time of the day and ambient light intensity, the system enters 'Night Mode'. In this mode, home heating and lighting are switched on. When the specified time/event triggers have occurred, the home server software accesses the home database for settings attached to that mode. It then switches the home to the required mode by changing the settings and states of each home device (logical or physical) to match that specified in the given mode.

## VII. SYSTEM TESTING AND EVALUATION

### A. Evaluation of System/GUI Software

A model of the system was developed in order to evaluate the viability of the proposed design. An ED, AP and wireless camera as implemented above were used for this evaluation. A lamp was connected to the expansion board. A heater was simulated by an on-board relay and LED. In addition, the camera control mechanism was connected to the ED. The AP was connected to the server. Then each feature of the system, as described above, was evaluated on a local client and server. Some of the results are shown in TABLE I.

TABLE I
RESULTS OF SYSTEM EVALUATION

| Component | Procedure | Expected Results | Actual Results |
|---|---|---|---|
| Temperature sensor | Brief application of hot air using soldering machine | Temperature rise | Temperature rose from 28 to 65 °C |
| Light sensor | -Covering of light sensor | -Decrease in light intensity reading | -Light reading dropped from 88 % to 8 % |
| | -Application of very bright light source close to sensor | -Increase in light reading | -Light reading increased from 88 % to 96 % |
| Motion sensor | Walking past the end device | Motion detection alert message | Detection indicated for 3 s |
| Lighting | -Switched by clicking ON/OFF GUI button | -Lamp should come on/off | -Lamp switched as required |
| | -Varied intensity using slider | -Light brighter/dimmer | -Light was made brighter/dimmer |
| Light/heater timer | The time that light/heater is switched on/off was set | Light/heater should come on/off at set time | Light/heater switched at expected time |
| Low/high threshold alarms | The low threshold value of light was set to 55%, low alarm action was set to "Turn light on" | Lighting should come on when light intensity is lowered | Lighting came on as required |
| Camera control | Clicked buttons to pan/tilt the camera | Camera should be panned tilted up/down | Camera moved as expected, total pan angle = 360°, tilt angle = 303°, Camera shakes slightly |

In general, overall system performance was adequate. The motion sensor indicated detection for up to 3 seconds after the event took place. The system could be made not to regard these as unique detections. Total camera tilt angle was limited to 303° by the mounting mechanism. To prevent damage to the camera and its control circuitry, users could be alerted when this limit is reached.

*B. Current and Voltage Measurements*

The current consumed by some components of the ED was determined. In addition, the voltages across these components were measured (TABLE II). In doing this, the ED was set up by connecting the camera motors and an LED lamp to it.

TABLE II
CURRENT AND VOLTAGE MEASUREMENTS OF END DEVICE COMPONENTS

| Component | Measured Current | Voltage (V) | Test Conditions |
|---|---|---|---|
| Temperature sensor | 2.26 µA | 2.9492 | 23 °C |
| Motion sensor | 1.90 µA | 2.9121 | - |
| Light sensor | 2.16 µA | 0.5323 | 260 Lux |
| End device | 23.50 mA | 3.0430 | - |
| Expansion board | 2.10 A | 5.0708 | |

The current consumption of the temperature, motion and light sensors (TABLE II) were within the maximum operating range specified in the relevant datasheets. The total current consumption of the ED (23.50 mA) is mostly used by the radio. Therefore, it is important to minimize the length of time the radio is active by putting it on sleep mode for most times [9]. When there is a message, the radio could be put back to active mode in order to receive the message. The high current consumption of the actuator driver circuit (2.10 A) was due to the high current rating of the two camera control motors (0.86 A each) connected to it. In order to reduce this current, smaller motors may be used at the expense of reduced power capacity. On the whole, the current and voltage readings were well within the recommended operating conditions, and validated the ED design.

*C. Range Testing*

The range of the system was tested by measuring the maximum distance between two wireless nodes in which there was reception of valid data. The test environment was the corridor on the third floor of Chesham Building (Wing B3), University of Bradford, Bradford, UK. There were two wooden doors with glass interiors in the test area to simulate walls in a home. The nodes were placed at a height of 0.4 m.

One node with an expansion board was kept stationary and periodically transmits sensor readings. The other node was moved till the point was reached where any further movement will lead to loss of valid data reception from the stationary node. Successful reception of frames was indicated by the blinking of an LED on the moving node. In addition, received signal strength indicator (RSSI) values reported by the moving AP node at various distances were recorded and are shown graphically in Fig. 7.

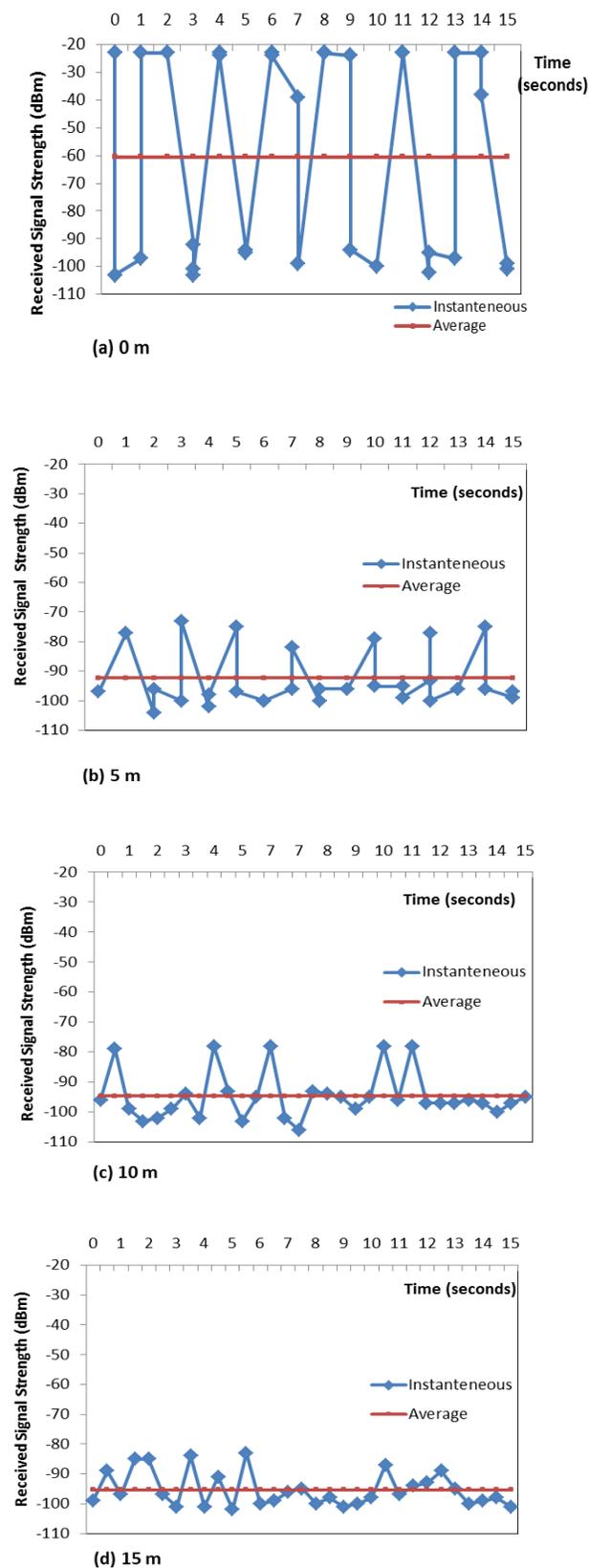

Fig. 7. RSSI values at various distances.

The maximum range was found to be 38.1 m, compared to the expected 50 m line-of-sight range specification of the wireless module. This could be due to the two doors in between the nodes, with maximum thickness of 12 cm and 15.5 cm. However, the observed range is sufficient for the use of the system in most homes. Where more range is needed, some nodes could be configured as repeaters.

From Fig. 7, the average RSSI value was -60.45 dBm at 0 m (the nodes are as close as possible), and dropped to -92.23 dBm at a distance of 5 m. The average RSSI value obtained at 10 m (-94.61 dBm) was close to that measured at 15 m (-95.29 dBm). Therefore, as expected, the average RSSI values roughly decrease with increase in distance [3].

However, it can be observed that at 15 m, some of the RSSI values were as low as -101 dBm. Such low values of RSSI could be due to multipath in the environment [3], [14], [15]. Signals reflected off walls in the environment, and weaker in strength, are received by the given node in addition to the original transmitted signal. Because the maximum sensitivity of the radio is -104 dBm, this leads to corrupted packets that are ignored by the receiving node, if detected at all. This could affect data integrity and reliability of sensor data or device control messages, leading to erratic system operation. However, there are mechanisms to reduce the effects of these problems, such as auto-acknowledgements and the use of transaction ID to detect duplicate messages. Frequency agility could help to reduce interference from other devices in the same frequency band.

## VIII. Conclusion

This work has presented the design and implementation of a home automation system featuring wireless sensors/actuator nodes and camera. It has also detailed the developed embedded and GUI software used to monitor and control the system. Integration with the internet also allowed remote access to the features of the home. With the use of modes based on explicit user requests, timers or environmental events, a layer of adaptive functionality was added to the WHAN. The developed system is easily extensible to accommodate new sensors and devices because of its modularity. Testing confirmed that the system's range was adequate for most home applications. Whereas the system was subjected to adverse multipath problems, suggestions to improve this situation were proffered. On the whole, the system provides a robust and cost-effective way of wirelessly monitoring and controlling the home environment and devices.